\documentclass[9pt,twocolumn,twoside]{osajnl}

\journal{ol} 

\setboolean{shortarticle}{true}

\title{Dark solitons under higher order dispersion}

\author[1*]{Tristram J. Alexander}
\author[2]{G. A. Tsolias}
\author[3]{A. Demirkaya}
\author[3]{Robert J. Decker}
\author[1]{C. Martijn de Sterke}
\author[2]{P. G. Kevrekidis}

\affil[1]{Institute of Photonics and Optical Science (IPOS), School of Physics, The University of Sydney, NSW 2006, Australia}
\affil[2]{Department of Mathematics and Statistics, University of Massachusetts, Amherst, MA 01003-4515, United States of America}
\affil[3]{Mathematics Department, University of Hartford, 200 Bloomfield Ave., West Hartford, CT 06117, United States of America}

\affil[*]{Corresponding author: tristram.alexander@sydney.edu.au}

\begin{abstract}
We show theoretically that dark solitons can exist in the presence of pure quartic dispersion, and also in the presence of both quadratic and quartic dispersive effects, displaying a much greater variety of possible solutions and dynamics than for pure quadratic dispersion. The interplay of the two dispersion orders may lead to oscillatory non-vanishing tails, which enables the possibility of bound, potentially stable, multi-soliton states. Dark soliton-like states which connect to low amplitude oscillations are also shown to be possible.  
Dynamical evolution results corroborate the stability picture obtained, and possible avenues for dark soliton generation are explored.
\end{abstract}

\setboolean{displaycopyright}{true}

\begin{document}

\maketitle

{\it Introduction.} Following the experimental observation of bright solitons supported by fourth-order dispersion~\cite{BlancoRedondoNC2016}, the theory of bright solitons with pure-quartic dispersion~\cite{TamOL2019}, mixed second- and fourth-order dispersion~\cite{TamPRA2020} and pure dispersion of higher even orders~\cite{RungePRR2021} has become well established. The experimental work was enabled by the development of a fiber laser with conveniently programmable dispersion \cite{RungePRR2021, RungeNP2020}. All of these bright solitons exist when the highest-order of dispersion is negative.  Concurrently, this has inspired substantial mathematical interest in the analysis of such models~\cite{beam_demirkaya,atanas,aceves,BandaraPRA2021}.

To the best of our knowledge, solitons with normal quartic dispersion, i.e., dark quartic solitons, have not been investigated.  Dark solitons appear as dark intensity dips on a continuous wave (CW) background~\cite{Emplit_1987,WeinerPRL1988,djf}.  Associated with the intensity minimum is a phase change of $\pi$, and this ``kink" in the phase provides the dark soliton with added (topological) robustness against noise~\cite{ZhaoOL1989}.  Recently, methods using self-induced modulational instability with normal dispersion have allowed the generation of dark soliton trains in optical fiber cavities~\cite{SongOL2014}, with evidence that dark solitons are ubiquitous in these conditions~\cite{TangPRA2013}. Indicative of their broader relevance is their emergence in both atomic systems~\cite{djf,siambook}, and in water waves~\cite{Chabchoub_2012,PhysRevLett.120.224102}.

The effect of dominant higher-order normal dispersion was considered in the $\phi^4$ Klein-Gordon model, in both the pure quartic form \cite{DeckerJPA2020,DeckerCNSNS2021} and in the presence of both quadratic and quartic terms \cite{TsoliasJPA2021}.  These results for real fields point to intricate soliton interactions.  
Dark solitons bear a non-zero background, unlike bright solitons, 
and, as we show, the quartic dispersion leads to oscillations in their non-vanishing tails.  This leads to a complex variety of effects absent in the purely quadratic case, accompanied by rich instability dynamics.


In this work we consider the fourth-order normal dispersion regime, in the presence of both normal and anomalous second-order dispersion and cubic Kerr nonlinearity. 
In the pure quadratic case there is a single dark soliton family of stationary solutions, with multiple dark solitons repelling one another. We find many more possibilities in the mixed quartic-quadratic dispersion case, opening up new directions for possible dark soliton experiments. We identify families of (multi-)dark solitons, examine their stability and instability numerically and dynamically, and conclude with a demonstration of possible dark soliton generation. 

{\it Model \& Theoretical Background.} 
We use the generalized nonlinear Schr\"{o}dinger equation for the electric field envelope $\tilde{\Psi}$ with quadratic and quartic dispersion and a Kerr nonlinearity:
\begin{equation}
    \label{model}
    i\frac{\partial\tilde{\Psi}}{\partial \xi} + \frac{\tilde{\beta}_4}{24}\frac{\partial^4\tilde{\Psi}}{\partial \tau^4} - \frac{\tilde{\beta}_2}{2} \frac{\partial^2\tilde{\Psi}}{\partial \tau^2} + \gamma|\tilde{\Psi}|^2\tilde{\Psi} = 0
\end{equation}
where $\xi$ is the propagation distance, $\tau$ is the retarded time in the frame of the pulse and $\gamma$ is the nonlinear coefficient, which we shall take to be positive. The parameters $\tilde{\beta}_2 = dv_g^{-1}/d\omega$ and $\tilde{\beta}_4=d^3v_g^{-1}/d\omega^3$, where $v_g$ is the group velocity and $\omega$ is the pulse carrier frequency, characterize the quadratic and quartic dispersion strengths respectively. We normalize the retarded time in units of $t_0 = 1$ ps, $t = \tau/t_0$, and similarly  the propagation length in terms of a characteristic propagation length $z_0 = 1$ mm, $z = \xi/z_0$.  Finally, we rescale $\Psi = \sqrt{z_0\gamma}\tilde{\Psi}$.  We restrict ourselves to positive (normal) quartic dispersion, and consequently fix $\tilde{\beta}_4 = 1$ ps$^4$ mm$^{-1}$, consistent with experiment~\cite{BlancoRedondoNC2016}, leading to $\beta_4 = +1$.  The resulting normalized model takes the form:
\begin{equation}
    \label{normalised}
    i\frac{\partial\Psi}{\partial z} + \frac{1}{24}\frac{\partial^4\Psi}{\partial t^4} - \frac{\beta_2}{2} \frac{\partial^2\Psi}{\partial t^2} + |\Psi|^2\Psi = 0,
\end{equation}
where the normalized quadratic dispersion parameter is given by $\beta_2 = \tilde{\beta}_2/(z_0t_0^2)$.  We look for stationary solutions $\Psi(t,z) = \psi(t)\exp(i\mu z)$, where $\mu$ is a nonlinearity-induced phase shift characterizing the stationary solution, leading to the following equation for the real (without loss of generality) amplitude $\psi(t)$:
\begin{equation}
    \label{stationary}
    -\mu\psi + \frac{1}{24}\frac{d^4\psi}{d t^4} - \frac{\beta_2}{2} \frac{d^2\psi}{d t^2} + \psi^3 = 0.
\end{equation}

We investigate dark soliton solutions to Eq.~(\ref{stationary}), 
connecting to the continuous wave background $\psi_{cw} \equiv \pm\sqrt{\mu}$ (which implies: $\mu > 0$). In the four dimensional phase space characterizing solutions to Eq.~(\ref{stationary}) these dark soliton solutions may either connect the positive CW solution back to itself (homoclinic solutions, for even numbers of solitons), or connect CW solutions of different signs (heteroclinic solutions, for odd numbers of solitons). We apply periodic boundary conditions so we only find homoclinic solutions.  We characterize these using the complementary power, which for a cavity of length $2L$ is $Q_c = \int_{-L}^{L} (\psi_{cw}^2 - \psi^2) dt$.

\begin{figure}
    \centering
    \includegraphics[width=84mm,clip = true]{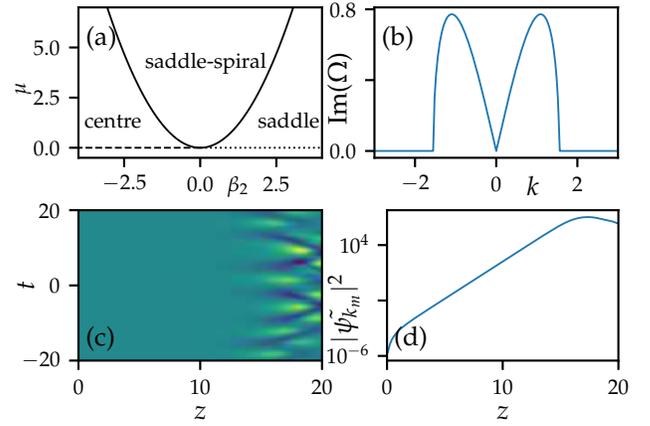}
    \caption{(a) Classification of CW solution with $\beta_2$ and $\mu$; (b) Modulational instability spectrum of CW for $\beta_2 = -0.2$, with most unstable wavenumber $k_m = 1.1$ corresponding to $\Omega = 0.77i$; (c) Instability dynamics of CW for $\beta_2 = -0.2$; and (d) Dependence of spectral intensity $|\tilde{\psi}|^2$ on time at wavenumber $k_m$ for results in (c), with slope giving $\Omega = 0.77i$, agreeing very well with the prediction.}
    \label{fig1}
\end{figure}

As a starting point, we analyze the nature of the CW, on top of which the dark solitons are built.
We consider purely real perturbations in the form $\psi = [\psi_{0} + \epsilon\exp(\lambda t)]$, and upon substitution of this perturbation into Eq.~(\ref{stationary}) and keeping only linear terms in $\epsilon$ we obtain the following equation for the perturbation eigenvalue $\lambda$:
\begin{equation}
    \label{zeroeig}
    \frac{1}{24}\lambda^4 - \frac{\beta_2}{2}\lambda^2 - \mu + 3\psi_{0}^2  = 0.
\end{equation}
Setting $\psi_0 = \psi_{cw}$ and solving for $\lambda$ we find that if $\mu\leq 3\beta_2^2/4$ we have four imaginary values for $\lambda$ if $\beta_2 < 0$, or four real values if $\beta_2> 0$.  The former implies that the CW is a center, so it is not possible to approach the CW and therefore no dark solitons connecting to the CW can exist.  Conversely, when $\lambda$ is purely real, the CW is a saddle point and we have the more familiar dark soliton regime.  If instead $\mu > 3\beta_2^2/4$ the $\lambda$ form a complex quartet, so the CW is a saddle-spiral: any approach to the CW is accompanied by oscillations. This behaviour is analogous to that observed for the bright soliton (where instead $\beta_4 < 0$). The dependence of the CW solution on the 2D parameter plane is shown in Fig.~\ref{fig1}(a); the quadratic dependence of $\mu$ on $\beta_2$ (solid curve), is as expected from the model's scaling properties.

As we shall see when we look at possible dark soliton solutions, the nature of the zero solution can also play a role in the form of the stationary states.  Substituting $\psi_0 = 0$ into Eq.~(\ref{zeroeig}) the eigenvalues $\lambda$ are an imaginary pair and a real pair, which makes the zero solution a saddle-center.  It is therefore not possible to approach zero asymptotically when $\beta_4 > 0$, but it is possible to connect to an oscillatory solution about zero.  This behaviour is distinct from the case of $\beta_4 < 0$ \cite{TamOL2019,TamPRA2020,BandaraPRA2021} where the zero solution is either approached asymptotically, or not at all.

While the analysis of the eigenvalues (\ref{zeroeig}) indicates where we can expect homoclinic (multi-)dark soliton solutions, a necessary condition for stability of these solutions is that the CW is modulationally stable.  To this end, we examine the growth of linear waves by more generally perturbing the CW background:
\begin{align}
    \label{pert}
    \Psi(t,z) = \big[\psi_{cw} &+ \epsilon_1\exp(ikt)\exp(i\Omega z)\\ \nonumber
   &+ \epsilon_2^*\exp(-ikt)\exp(-i\Omega z)\big] \exp(i\mu z), 
\end{align}
where the $\epsilon_i$ denote small perturbations.  Substitution of (\ref{pert}) into 
Eq.~(\ref{normalised}) 
shows how the perturbation eigenvalue $\Omega$ is related to the perturbation wavenumber $k$:
\begin{equation}
    \label{det}
    \Omega^2 = \left(-\mu+\frac{\beta_2}{2}k^2+\frac{1}{24}k^4+2 \psi_{cw}^2\right)^2 -  \psi_{cw}^4.
\end{equation}
If $\Omega^2 < 0$ then the perturbation undergoes exponential growth, with the boundary of modulational stability/instability occurring when $\Omega^2 = 0$. We see that we always have an acoustic branch of modulational instability if $\beta_2 < 0$ (see Fig.~\ref{fig1}(b) for $\beta_2 = -0.2$).  Instability in this region proceeds with characteristic modulation of the background (Fig.~\ref{fig1}(c)) and exponential growth of the most unstable wavenumber $k_m$ (Fig.~\ref{fig1}(d)).  We see almost perfect agreement between the growth rate predicted by Eq.~(\ref{det}) and the numerically observed value.  The CW is modulationally stable if $\beta_2 \ge 0$ and $\mu > 0$.
\begin{figure}
    \centering
    \includegraphics[width=84mm,clip = true]{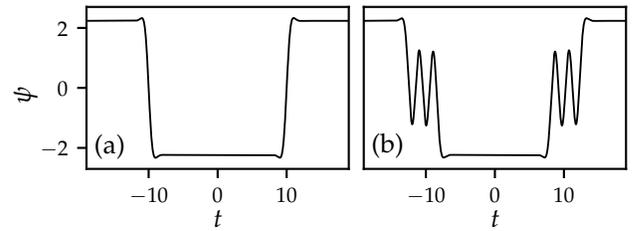}
    \caption{Pure quartic dark soliton stationary solutions ($\beta_2 = 0$, $\mu = 5$). (a) A well separated pair of dark solitons; (b) A mixed solution connecting dark solitons to an oscillation about $0$.}
    \label{fig2}
\end{figure}

{\it Numerical Findings} All propagation results (including the CW dynamics in Fig.~\ref{fig1}) are obtained using a fourth-order split-step numerical scheme~\cite{Yang2010} applied to Eq.~(\ref{normalised}).  Here, $\Delta t = 9.8 \times 10^{-3}$ with $\Delta z = 7.6 \times 10^{-6}$. The robustness of the scheme is monitored by evaluating the Hamiltonian $H = \int|d^2\psi/dt^2|^2+\beta_2|d\psi/dt|^2+|\psi|^4 dt$ and verifying that this quantity is conserved during propagation.  Stationary solutions are found by numerically solving Eq.~(\ref{stationary}) using a conjugate gradient method~\cite{Yang2010}.

The linear analysis points to new possible features in dark soliton solutions. Figure~\ref{fig2} shows the two extremes of possible solutions in the pure-quartic case.  Well separated dark solitons can be found (Fig.~\ref{fig2}(a)), similar to those observed in the quadratic case, but with characteristic damped oscillations approaching the CW, consistent with earlier results for the real case \cite{DeckerCNSNS2021}.  Fig.~\ref{fig2}(b) shows a new possibility enabled by the quartic dispersion, a connection between the CW and the saddle-centre at the origin.
\begin{figure}
    \centering
    \includegraphics[width=84mm]{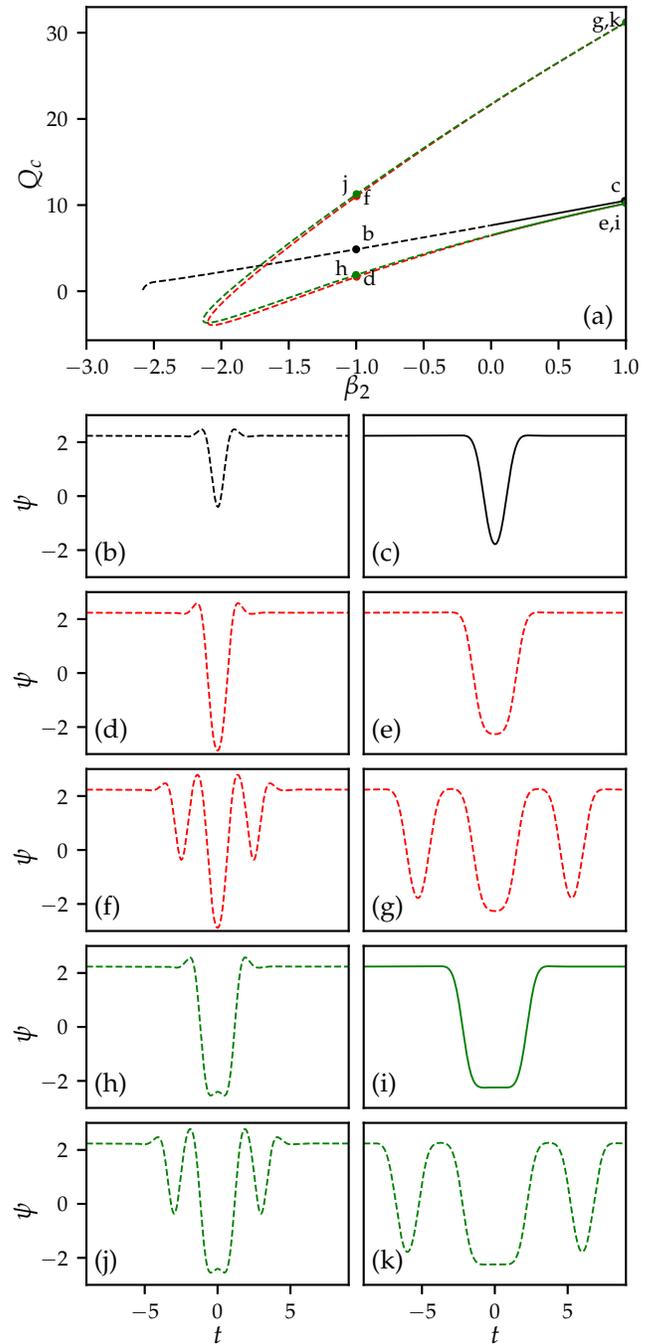}
    \caption{(a) Bifurcation diagram for the lowest order dark soliton families with the labels corresponding to solutions shown in the lower panels. (b) and (c) solutions corresponding to family 0, ultimately connecting to a plane wave as $\beta_2$ decreases; (d),(e),(f) and (g) are solutions from family 1; (h),(i),(j) and (k) solutions from family 2.  Left and right panels correspond to $\beta_2 = -1$, $\beta_2 = 1$, respectively.  Unstable/stable solutions shown with dashed/solid lines respectively. In all cases $\mu = 5$.}
    \label{fig3}
\end{figure}

Examining the possible solutions more systematically as a function of a system parameter, e.g. $\beta_2$ in Fig.~\ref{fig3}(a), we find that many possible families exist, each with a characteristic spacing between the dark soliton pairs.  The most closely spaced dark soliton pairs, which we call `family $0$' (Figs.~\ref{fig3}(b) and (c)), connect the CW to the origin, and we find bifurcate from the CW solution at just less than $\beta_2 = -2.5$.  In contrast, all other families of increasingly widely spaced dark soliton pairs (e.g. `family $1$' and `family $2$' in Figs.~\ref{fig3}(e) and (i) respectively), develop additional undulations as they proceed to negative $\beta_2$ (Figs.~\ref{fig3}(d) and (h)), where they collide (in a saddle-center bifurcation) with an upper branch associated with large side oscillations (Fig.~\ref{fig3}(f) and (j)).  These upper branch solutions have additional dark soliton pairs on either side of the main pair, i.e., they are a
triple pair family, which becomes increasingly evident as $\beta_2$ increases (Fig.~\ref{fig3}(g) and (k)).  These results are as expected from the earlier real analysis, with the oscillatory tails enabling isolated distances at which dark solitons become stationary.   

We examine the stability of the solutions shown in Fig.~\ref{fig3} by perturbing the stationary solution and numerically computing the linearization  eigenvalues~\cite{Yang2010}.  A full presentation of the linear stability analysis is beyond the scope of this work, however we find results consistent with the real case~\cite{DeckerCNSNS2021}.  We find that family $0$ and family $2$ are stable for $\beta_2 \ge 0$ (Fig.~\ref{fig3}(c)), so stability alternates with spacing.  This is topologically justified from the underlying alternating minima and maxima in the relevant effective landscape (c.f., e.g., Fig.~5 of~\cite{TsoliasJPA2021} in the real-field case). All solutions are unstable for $\beta_2 < 0$ due to the modulational instability of the CW background.  The continuous spectrum, as is typical for NLS dark solitons, spans the entire imaginary axis~\cite{siambook}.
Stable and unstable families of solutions are shown respectively as solid and dashed lines in Fig.~\ref{fig3}.
\begin{figure}
    \centering
    \includegraphics[width=84mm,clip = true]{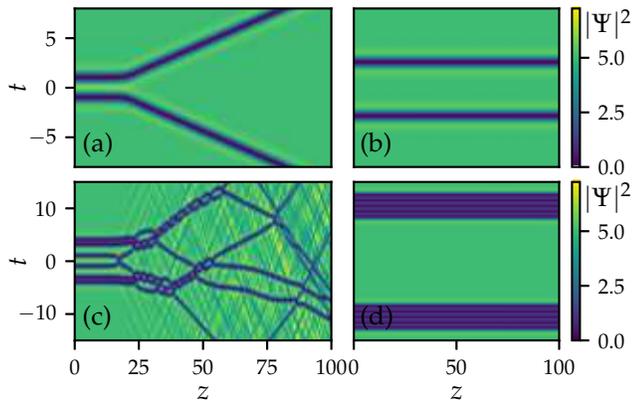}
    \caption{Numerical propagation of dark soliton solutions from different families at $\beta_2 = 0$. (a) Family 1 is unstable to fission; (b) Family 2 is stable for $\beta_2 > 0$, due to inversion of the stability one step further apart; (c) Family 2 upper branch, all upper branch composite solutions are unstable; (d) Solution shown in Fig.~\ref{fig2}(b), a very weak oscillatory instability is present.}
    \label{fig4}
\end{figure}

Examples of possible stability and instability dynamics are shown in Fig.~\ref{fig4} for $\beta_2 = 0$.  Figs.~\ref{fig4}(a) and (b) confirm respectively the instability of family $1$ and the stability of family $2$. While all upper branch solutions are unstable, the instability dynamics show long-lived non-stationary soliton dynamics which, as we shall see, appear to play a significant role in dark soliton generation processes.  The instability dynamics shown in Fig.~\ref{fig4}(c) are one example, with more closely bound pairs appearing robust until disturbed by the instability products of the inner pair.  The generalisation of family $0$ to multiple oscillations about the zero solution, corresponding to Fig.~\ref{fig2}(d), appears remarkably robust, as seen in Fig.~\ref{fig4}(d), but stability analysis reveals a weak oscillatory instability.

\begin{figure}
    \centering
    \includegraphics[width=84mm,clip = true]{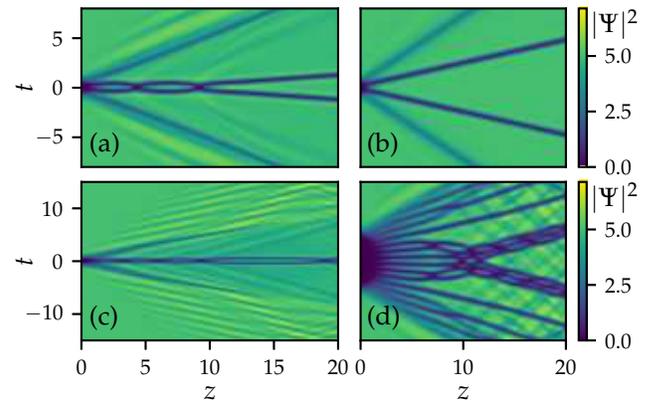}
    \caption{Dark soliton generation from intensity notches in a CW background.  (a,b,c) Initial condition $\psi(t,0) = \sqrt{\mu}(1-\exp(-t^2/2))$, (d) initial condition $\psi(t,0) = \sqrt{\mu}(1+(\tanh(t-5)-\tanh(t+5))/2)$, all with $\mu = 5$; (a) Pure quartic dark soliton bound state with $\beta_2 = 0$; (b) Repulsive dark soliton pair ($\beta_2 = 0.5$); (c) Dark soliton pair when the CW is modulationally unstable ($\beta_2 = -0.2$); (d) Dark soliton complex at $\beta_2 = 0$.}
    \label{fig5}
\end{figure}

Given the complexities of controlling dispersion (e.g., using a WaveShaper) in the bright soliton case\cite{RungeNP2020,RungePRR2021}, the careful pulse control of early dark soliton generation experiments~\cite{WeinerPRL1988} appears to be practically challenging. We therefore restrict considerations to an, arguably, prototypical initial condition of a CW state with intensity notches, either of Gaussian form or domain wall-like form using a hyperbolic tangent.
With a Gaussian initial condition we consider the dynamics in three different regimes: the pure quartic case ($\beta_2 = 0$, Fig.~\ref{fig5}(a)), positive $\beta_2$ ($\beta_2 = 0.5$, Fig.~\ref{fig5}(b)), and negative $\beta_2$ ($\beta_2 = -0.2$, Fig.~\ref{fig5}(c)).  In the pure-quartic case a dark soliton pair appears, and exhibits a periodic cycle of collisions and separations, before escaping the local potential induced through the oscillatory tails, whereas for the positive $\beta_2$ example, the dark solitons repel each other (which we note is the only possible dynamics in the pure positive quadratic dispersion case).  For the negative $\beta_2$ case the dark soliton pair is more tightly bound, but sits on a modulationally unstable CW background. A domain wall-like initial condition produces multiple dark soliton pairs, and even what appears to be a travelling complex of dark solitons, Fig.~\ref{fig5}(d), similar to the instability dynamics observed earlier.  We see that in the presence of instability (soliton or CW), solitons emerge, although they may be moving solitons.  The nature of these moving solitons is an interesting direction for further research, particularly in light of recent results indicating the non-Galilean invariance of solitons in the presence of quartic dispersion~\cite{WidjajaPRA2021}.

Typical experimental values for associated optical settings would involve a nonlinearity coefficient of $\gamma = 4.07$ W$^{-1}$mm$^{-1}$~\cite{BlancoRedondoNC2016}.  In this case, taking the temporal units to be $1~{\rm ps}$ and the propagation units to be $1~{\rm mm}$, the power of the CW considered throughout is $1.2~{\rm W}$, the pulse length is $40~{\rm ps}$, and the propagation distance is $100~{\rm mm}$. These parameters can be adjusted via the scaling properties of the model.

{\it Conclusions \& Outlook.} 
This work provides the first study of higher-order dispersion dark solitons in an optical context, focusing on soliton families and their properties in pure quartic and quadratic/quartic dispersion settings. A first example of a mixed state dark soliton is given, in which the CW is connected to oscillations about the zero solution.  
The complex bifurcation picture uncovered for dark soliton pairs requires further study, including examining the properties of the waveforms as the modulationally unstable regime is approached.  The challenge of generating these states indicates a wealth of phenomena that could be further explored, including soliton complexes and gray solitons.  Our simulations indicate that even for the simple unstable states, it appears possible to observe long-lived interacting states in the dynamics.  We find no stable stationary states in the presence of positive $\beta_4$ and negative $\beta_2$, raising the question of the nature of the field in this region of parameter space. 


\bibliography{Dark}

\end{document}